\documentclass[aps,prl,twocolumn,superscriptaddress]{revtex4-1}
\usepackage{amsmath}
\usepackage[pdftex]{graphicx}
\usepackage[thinspace]{SIunits}

\providecommand \BibitemShut  [1]{\csname bibitem#1\endcsname}%

\begin{document}

%\preprint{}

\title{Electrical control of inter-dot electron tunneling in a quantum dot molecule}

\author{K. M\"uller}
 \affiliation{Walter Schottky Institut and Physik-Department, Technische Universit\"at M\"unchen, Am Coulombwall 4, 85748 Garching, Germany \\}
\author{A. Bechtold}
 \affiliation{Walter Schottky Institut and Physik-Department, Technische Universit\"at M\"unchen, Am Coulombwall 4, 85748 Garching, Germany \\}
\author{C. Ruppert}
 \affiliation{Experimentelle Physik 2, TU Dortmund, 44221 Dortmund, Germany \\}
\author{M. Zecherle}
 \affiliation{Physik-Department E11, Technische Universit\"at M\"unchen, James-Franck-Str, 85748 Garching, Germany \\}
\author{G. Reithmaier}
 \affiliation{Walter Schottky Institut and Physik-Department, Technische Universit\"at M\"unchen, Am Coulombwall 4, 85748 Garching, Germany \\}
\author{M. Bichler}
 \affiliation{Walter Schottky Institut and Physik-Department, Technische Universit\"at M\"unchen, Am Coulombwall 4, 85748 Garching, Germany \\}
\author{H. J. Krenner}
 \affiliation{Lehrstuhl f\"ur Experimentalphysik 1 and Augsburg Centre for Innovative Technologies (ACIT), Universit\"at Augsburg, Universit\"atsstr. 1,
86159 Augsburg, Germany\\}
\author{G. Abstreiter}
 \affiliation{Walter Schottky Institut and Physik-Department, Technische Universit\"at M\"unchen, Am Coulombwall 4, 85748 Garching, Germany \\}
\author{A. W. Holleitner}
 \affiliation{Walter Schottky Institut and Physik-Department, Technische Universit\"at M\"unchen, Am Coulombwall 4, 85748 Garching, Germany \\}
\author{J. M. Villas-Boas}
 \affiliation{Instituto de F\'isica, Universidade Federal de Uberl\^andia, 38400-902 Uberl\^andia, MG, Brazil \\}
\author{M. Betz}
 \affiliation{Experimentelle Physik 2, TU Dortmund, 44221 Dortmund, Germany \\}
\author{J.J. Finley}
 \email{finley@wsi.tum.de}
 \affiliation{Walter Schottky Institut and Physik-Department, Technische Universit\"at M\"unchen, Am Coulombwall 4, 85748 Garching, Germany \\}

\date{\today}

\begin{abstract}
We employ ultrafast pump-probe spectroscopy to  directly monitor electron tunneling between discrete orbital states in a pair of spatially separated quantum dots.  Immediately after excitation, several peaks are observed in the pump-probe spectrum due to  Coulomb interactions between the photo-generated charge carriers. By tuning the relative energy of the orbital states in the two dots and monitoring the temporal evolution of the pump-probe spectra the electron and hole tunneling times are separately measured and resonant tunneling between the two dots is shown to be mediated both by elastic and inelastic processes.  Ultrafast ($< 5 \, ps$) inter-dot tunneling is shown to occur over a surprisingly wide bandwidth, up to $\sim8meV$, reflecting the spectrum of exciton-acoustic phonon coupling in the system.
\end{abstract}

%\keywords{}

\maketitle

% \% Put \label in argument of \section for cross-referencing

%TEXT FOR INTRODUCTION
Tunneling is a widespread phenomenon in both low- and high-energy physics governing the dynamics of nuclear decay, tunnel ionization of atoms and vertical electron transport in semiconductor heterostructures. Tunneling generally occurs to and/or from a \textit{continuous} distribution of initial or final quantum states associated with particles having one or more motional degrees of freedom. However, tunneling between discrete quantum states in solids having a specific energy separation is of particular interest since additional quasiparticles may be required to ensure energy conservation. Semiconductor nanostructures are an ideal test bed for such processes since energy levels can be arbitrarily designed whilst coupling to the phonon bath can facilitate inelastic tunneling. So far, inelastic tunneling between fully localized states has been studied only indirectly by transport measurements in electrostatically defined double quantum dots (QDs) \cite{PhysRevLett.86.878, Qin01} and in locally gated carbon nanotubes \cite{Mason30012004}.  In strong contrast, self-assembled QDs offer the advantage of strongly localized electron and hole states and the possibility to perform time-resolved \textit{optical} measurements to directly monitor carrier populations in real time. Vertically stacking of optically active QDs produces quantum dot molecules (QDMs) with strong, electrically tunable tunnel coupling giving rise to pronounced anticrossings in optical experiments \cite{PhysRevLett.94.057402, Stinaff2006, PhysRevLett.97.076403}. Tunnel coupling can occur for either electrons or holes \cite{Bracker2006}, depending on the relative dot size and couplings between different orbital states can occur due to symmetry breaking \cite{Scheibner2008}.

In this letter we report the direct observation of ultrafast electron transfer between discrete quantum states in a pair of spatially separated QDs with electrically tunable coupling.  Picosecond pump-probe spectroscopy is employed to monitor the evolution of the optical response of this ``molecular'' QD system over the first few hundred picoseconds after an exciton has been created. Immediately after excitation, a number of pronounced features appear due to pump-induced absorption and the resulting few-Fermion interactions in the system. By monitoring the temporal evolution of the pump-probe spectrum we independently track both electron and hole dynamics. As the electron quantum states in each dot are tuned through energetic resonance, we observe ultrafast inter-dot tunneling and, surprisingly, find that inelastic processes play a key role. Analysis of the results allows us to map out the spectral function of the exciton-phonon interaction, demonstrating that inter-dot tunnelling occurs over a surprisingly wide bandwidth, and within $< 5 \, ps$ at resonance.

The sample investigated consisted of a vertically stacked pair of self-assembled InGaAs QDs separated by a 10 nm thick GaAs spacer and embedded within the intrinsic region of an n-type GaAs Schottky photodiode \cite{Mueller11}. Complementary photocurrent (PC) absorption and photoluminescence (PL) emission measurements can be performed by varying the applied gate voltage. For the QDM investigated in this study we observe a strong anticrossing in PL.\cite{complex}  This arises from tunnel coupling of the electrons between the two dots forming the molecule \cite{PhysRevLett.94.057402, Mueller11}.  As shown in the field dependent PL spectra in Fig.1, the s-orbital states in the two dots are tuned into resonance at an electric field (F) of $F_0 = 23.1 \, kV/cm$ and couple with a strength $2V_0 = 3.4 \, meV$. This coupling is manifested by an anticrossing in the PL data of direct and indirect neutral exciton transitions with the hole  located in the upper QD and the electron occupying either the lowest energy orbital state in the upper dot ($e_{ud}$) or lower dot ($e_{ld}$), respectively (see ref \cite{Mueller11}). As the electric field increases the PL quenches, due to tunneling escape of electrons and holes from the dot, and the direct exciton in the upper dot can continue to be monitored in PC-experiments. As shown in the right part of  Fig.1, this direct exciton exhibits a second, much weaker anticrossing at $F_1=33.1 \, kV/cm$ with a much weaker coupling of $2V_1 = 0.8 \, meV << 2V_0$. We identify the second anticrossing as depicted schematically in the insets of Fig.1; the lowest electron level in the upper dot ($e_{ud}$ - left inset Fig.1) is tuned into resonance with an excited orbital electron state in the lower dot ($e_{ld}^*$ - right inset) \cite{parity}. The difference between the electric fields where the two anticrossings occur ($F_1 - F_0 = 10.0 \pm 0.1 \, kV/cm$) is a direct measure of the energy separation $\Delta E_{ld}^{2-1}$ of the two electron orbitals in the lower dot. Using the static dipole moment $ed = e \times 15.3 \pm 0.1 \, nm$\cite{Mueller11}, we obtain $\Delta E_{ld}^{2-1} = 15.3 \pm 0.25 \, meV$, fully consistent with the expected orbital energy spacings of the QD-molecule.  The results presented in Fig. 1 were found to be quite general, similar observations having been made for different samples grown under nominally identical conditions.

\begin{figure}
\includegraphics[width=1\columnwidth]{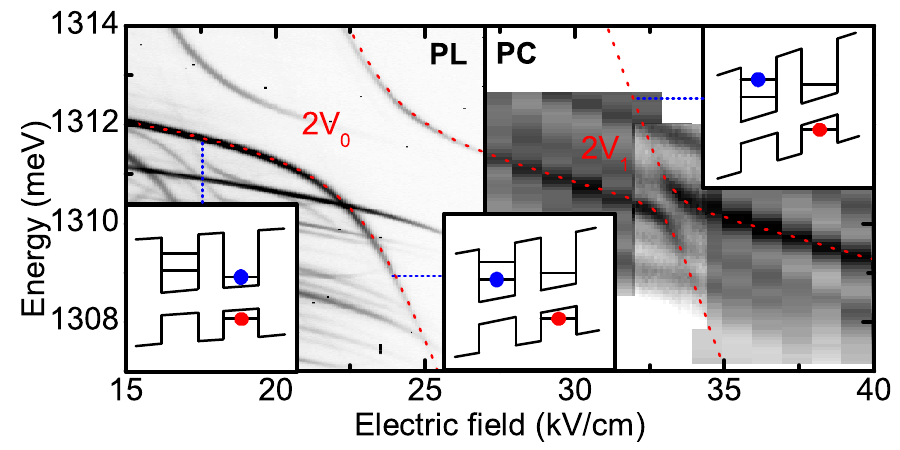}
\caption{\label{fig:Figure_1}
(Color online) Photoluminescence emission (left part) and photocurrent absorption spectra recorded at $T=10 \, K$ from the molecule. Anticrossings related to tunnel coupling of the electron in the ground state of the upper dot and the ground (or excited) state of the lower dot.}
\end{figure}

We continue by investigating the ultrafast dynamics of charge carriers in the QDM using pump-probe experiments with PC readout. The setup used (ref. \cite{Zecherle2010}) provides two, independently tunable  $3.8 \pm 1 \, ps$ duration laser pulses derived from a single femtosecond Ti:Sapphire source.  After identifying the discrete neutral exciton transition in the upper dot using CW spectroscopy \cite{PhysRevLett.94.057402,PhysRevLett.97.076403,Mueller11}, we selectively excite it with a resonant pump pulse.  The photocurrent $I$ induced by the time-delayed probe pulse is then measured with a lock-in amplifier. Recording the difference spectrum obtained with the pump beam blocked and unblocked, respectively, provides  the \textit{pump-induced} change of the PC signal, $\Delta I$. The quantities $I$ and $\Delta I$ can be interpreted as the absorption by the upper dot and its pump induced change. Typical examples of pump-probe PC spectra recorded at $F = 32.4 \, kV/cm$, away from the anticrossings, and three different time delays $t_D = 10, 50$ and $100 \, ps$ are presented in Fig.2(a). For these spectra the pump pulse was fixed to the direct neutral exciton in the upper dot $X$ at $1310.3 \, meV$  while the probe pulse was scanned. The figure clearly exhibits two positive going resonances (pump induced absorption) at $1308.3 \, meV$ and $1309.9 \, meV$, respectively.  In addition, a pronounced negative going dip (pump induced bleaching) is observed at the energy of $X$ at $1310.3 \, meV$. The observed pump-probe spectra can be understood with the level scheme depicted schematically in Fig.2(b) \cite{Zecherle2010, Ramsay08, Boyle08}. The pump pulse generates an exciton in the upper QD and, providing that the delay between pump and probe is less than the timescales for electron and hole tunneling, the probe pulse can further excite the system to generate the spatially direct biexciton. This explains the pronounced pump induced absorption peak $X \rightarrow 2X$ at $1308.3 \, meV$, red-shifted by $2 \, meV$ from the exciton transition $cgs \rightarrow X$. As the time delay increases, carriers tunnel out of the upper dot with the tunneling time of the electron ($t_e$) with its smaller effective mass being much shorter than that of the hole ($t_h$). If the electron tunnels out of the upper dot, leaving it occupied by a single hole the system exhibits an induced absorption peak at the $h \rightarrow X^+$ transition (peak at $1309.9 \, meV$).The anticorrelated intensity of $ X \rightarrow 2X$ and $h \rightarrow X^+$ can clearly be observed in Fig.2 as $t_D$ increases from 15 to 100 ps. If pump and probe pulse are both centered at the exciton transition $cgs \rightarrow X$, the probe pulse detects a bleaching of the exciton, resulting in the strong negative signal $\Delta I$ observed in Fig.2(a). 
Multi-peak Lorentzian fits to the transient photocurrent response (Fig. 2(a) full lines) yields linewidths of $670\pm50\mu eV$ for the $X \rightarrow 2X$ transition, $360\pm70\mu eV$ for $h^+ \rightarrow X^+$ and $340\pm70 \mu eV$ for $ cgs \rightarrow X$. When deconvoluted with the laser ($170\pm 40 \mu eV$), we obtain a ratio of the two-electron ($X \rightarrow 2X$) and one-electron ($h^+ \rightarrow X^+$, $ cgs \rightarrow X$) transition linewidths of $2.7\pm1.3$, an observation that is likely to be related to higher electron tunnelling probability for the biexciton final state that contains \textit{two} electrons, each of which can tunnel. In this very simplistic picture, one would expect the tunnelling lifetime of the biexciton to be approximately \textit{half} of the lifetime of either $X$ or $X^+$ and, hence, its linewidth to be $\sim3\times$ larger \cite{PhysRevB.73.125304} consistent with the observed linewidths.”

\begin{figure}
\includegraphics[width=1\columnwidth]{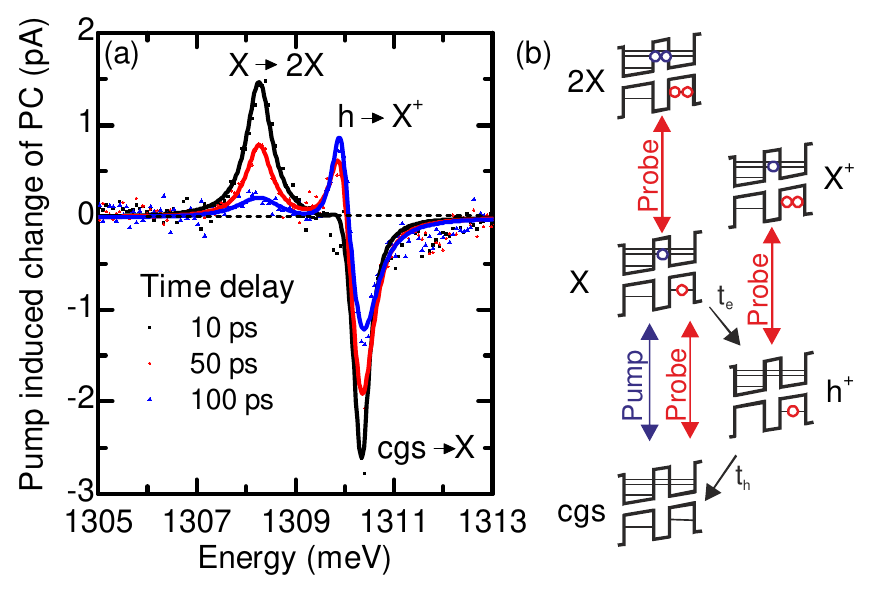}
\caption{\label{fig:Figure_2}
(Color online) (a) Pump induced change of the probe induced photocurrent for various time delays elapsed since excitation. (b) Level scheme for the pump-probe experiments on the QDM. }
\end{figure}

\begin{figure}
\includegraphics[width=1\columnwidth]{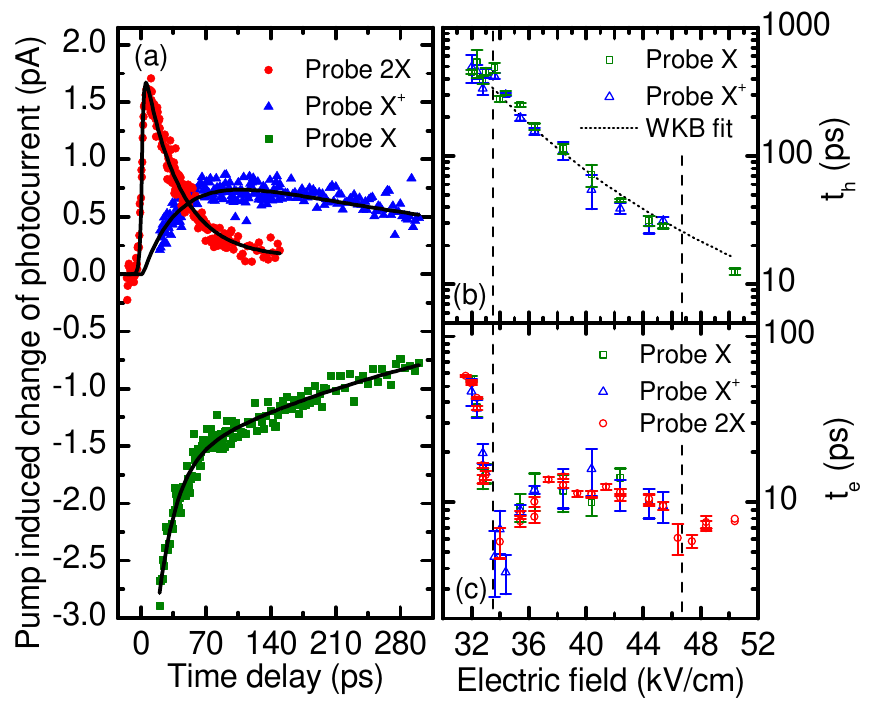}
\caption{\label{fig:Figure_3}
(Color online) (a) Temporal evolution of PC change $\Delta I$ probing the resonances identified in Fig.1(c). (b) Hole and (c) electron tunneling times $t_h$ and $t_e$ as a function of the applied electric field $F$.}
\end{figure}

To quantitatively analyze electron and hole tunneling, the pump photon energy is fixed to the $cgs \rightarrow X$ transition, while the probe photon energy is tuned to the three resonances identified in Fig.2(a). For each curve in Fig.3(a) $\Delta I$ is presented as a function of $t_D$, for probe pulses tuned to the $cgs \rightarrow X$ (green squares), $h^+ \rightarrow X^+$ (blue triangles) and $X \rightarrow 2X$ (red circles) transitions, respectively. Fits to each set of data using a rate equation model that accounts for sequential electron and hole tunneling \cite{Zecherle2010}(lines in Fig.3a) show excellent global agreement with all of the measured data and permit the direct, and independent, determination of $t_h$ and $t_e$. To investigate the influence of the coupling presented in Fig.1, detailed studies were performed over the range of electric field  $31 \, kV/cm \leq F \leq -51 \,kV/cm$ encompassing the anticrossing presented in Fig.1. The obtained values of $t_h$ obtained with the probe pulse tuned to the $cgs \rightarrow X$ (green squares) and the $h^+ \rightarrow X^+$ (blue triangles) transition are presented in Fig.3(b). Similarly, the field dependence of $t_e$ extracted by probing either the $cgs \rightarrow X$  (green squares - Fig.3c), $h^+ \rightarrow X^+$ (blue triangles)  and $X \rightarrow 2X$ (red circles) transition are plotted in Fig.3(c). In all cases, the tunneling times obtained by probing the different transitions are in excellent agreement with each other supporting the overall validity of the interpretation and the level scheme used (Fig.2b) over the whole range of $F$ analyzed here. The hole tunneling time monotonically decreases from $t_h = 450 \, ps$ at $F = 32 \, kV/cm$ to $t_h = 12.6 \, ps$ at $F = 50.5 \, kV/cm$ and exhibits a clear $~exp(-1/F)$ dependence. This behaviour is fully explained using the WKB tunelling rate \cite{Villas-Boas2005}
\begin{equation}
\Gamma = \frac{16V_s^2L}{\pi^2\hbar^2}\sqrt{\frac{m_s^*}{2|E_s|}}exp\left[ - \frac{4\sqrt{2m_s^*}}{3\hbar e F} |E_s|^{\frac{3}{2}}\right]
\label{WKB}
\end{equation}where $m_s^*$ is the effective mass of the particle s (electron or hole), $L$ and $V_s$ are the width and effective depth of the potential well and $E_s$ is the single particle quantization energy. A fit to the experimental data using eq. \ref{WKB} is presented as a dashed line in Fig.3(b) and consistently reproduces the experimental data. While the hole tunelling time is well accounted for by WKB theory unaffected by the lower QD, the field dependence of $t_e$ (Fig.3c) exhibits significantly richer behavior. We observe a series of resonances in $t_e$ marked with dashed lines on Fig.3(c).  Close to these resonances $t_e$ becomes comparable to the duration of the probe laser pulse. The most pronounced resonance occurs at $F_1 = 33.1 \, kV/cm$ precisely where the anticrossing presented in Fig.1 occurs. Here, the electron wavefunction delocalizes across the two dots forming the molecule and the electron tunnelling time is limited by the escape time from the excited orbital state in the lower dot ($e_{ld}^*$).  A second, weaker, dip is observed at $F_2 = 46.5 \, kV/cm$ where $e_{ud}$ anticrosses with a further excited orbital state of the electron in the lower dot $e_{ld}^{2*}$ (see Supplementary Material).

\begin{figure}
\includegraphics[width=1\columnwidth]{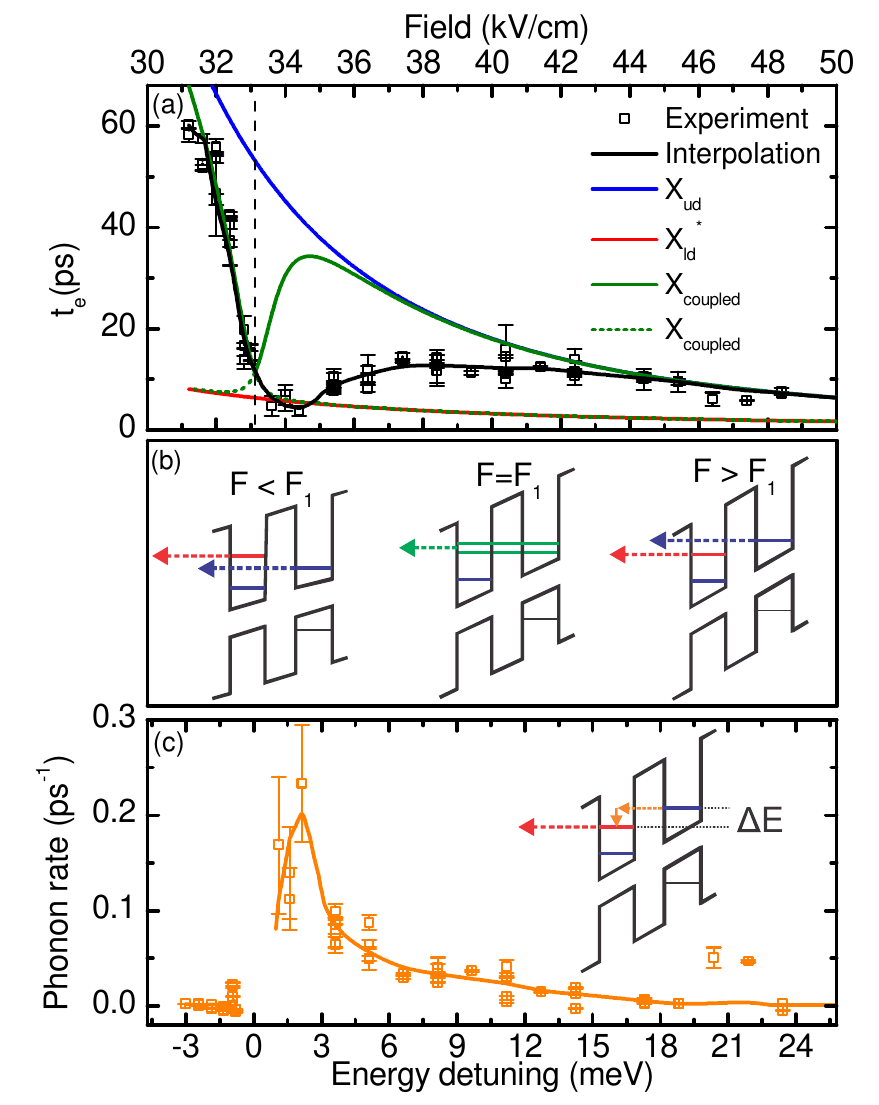}
\caption{\label{fig:Figure_4}
(Color online) (a) Tunneling time of the electron $t_e$. Measured data is shown as points. Theoretical analysis of the tunneling times of $e_{ud}$ ($e_{ld}^*$) is presented as a blue (red) line and for the coupled case as green lines. (b) Schematic illustration of the alignment and tunneling of $e_{ud}$ and $e_{ld}^*$
(C)Phonon rate extracted from (a)}
\end{figure}

We continue to present a detailed analysis of the ultrafast electron tunneling in the vicinity of the anticrossing at $F_1 = 33.1 \, kV/cm$ and show that it results from tunneling between the two dots forming the molecule with and without the participation of phonons. The electron tunneling time $t_e$ as a function of $F$ is plotted in Fig.4(a) on a linear scale, the position of the anticrossing is indicated by the vertical dashed line. In our theoretical analysis we consider tunnel processes according to eq. \ref{WKB} for the two electron levels involved in the anticrossing $e_{ud}$ and $e_{ld}^*$, as schematically depicted in blue and red in Fig.4(b). Since the two dots forming the QDM are similar in size and electronic structure, we assume identical confinement potentials for $e_{ud}$ and $e_{ld}$. As discussed above, $e_{ld}^*$ lies $15.3 \, meV$ to higher energy than $e_{ld}$, reducing the effective tunnel barrier. Thus, generically we expect electron tunneling out of the lower dot into the continuum to be faster from $e_{ld}^*$ compared with $e_{ld}$. When the two dots are coupled by tunneling (e.g. for $F = F_1$ - Fig.4(b)) the Hamiltonian is $\bf{H} = \bf{H}_0 + \bf{V}_1$, where $\bf{H}_0$ is the Hamiltonian of the detuned system and the coupling term $\bf{V}_1$ is measured from the PC data shown in Fig.1. The effective broadening of the coupled states is then given by the imaginary part of the eigenvalues of the coupled eigenstates $\epsilon_i=E_i-i\Gamma_i$
\begin{equation}
2\epsilon_\pm=\epsilon_{ld^*}+\epsilon_{ud}\pm\sqrt{(\epsilon_{ld^*}-\epsilon_{ud})^2+4V_1^2}
\label{Energy}
\end{equation}
Far from resonance the tunnel rates from $e_{ud}$ can be fitted using WKB formalism. This results in the blue (red) curve for $e_{ud}$ ($e_{ld}^*$) presented in Fig.4(a), taking only the measured energy difference between $e_{ld}$ and $e_{ld}^*$ of $15.3 \, meV$ into account. At resonance, the levels are mixed by tunneling and the decay rates of the coupled states exchange character with an intersection at $F_1$. The corresponding calculated values of $t_e$ are presented as green lines in Fig.4(a). For all electric fields we pump the state which is more located in the upper QD (solid green line), resulting in a dip of $t_e$ at $F_1$. The theoretical model quantitatively explains the measured values of $t_e$ for $F<F_1$ and $F>>F_1$. However, pure resonant tunneling would give rise to a much narrower resonance (green curve - Fig.4)  than is observed experimentally.  In experiment, the resonance is asymmetrically broadened for $F>F_1$. Closer inspection even shows that the shortest time $t_e$ is not observed at $F_1$ as expected from the theory but shifted by $\Delta F = +1 \, kV/cm$ to $34.1 \, kV/cm$. These clear experimental observations show that an additional mechanism must be active to shorten $t_e$ for $F_1<F< 41 \, kV/cm$ and produce the asymmetric resonance observed (Fig.4a). 

We continue to present evidence that this mechanism is acoustic phonon mediated inelastic inter-dot tunneling into $e_{ld}^*$. First, we compute the decay rate of an additional scattering mechanism from the difference between the measured tunneling times $t_e$ (black curve - Fig.4(a)) and the expected one from the model based on eq. \ref{Energy} (green curve). Taking into account the measured static dipole moment $ed = e \times 15.3 \pm 0.1 \, nm$  of the indirect exciton the additional tunneling rate can be obtained as a function of the energy detuning between the tunnel-coupled states. The result of this analysis is plotted in Fig.4(c) revealing a clear maximum for a positive detuning of $+1.7 \pm 0.1 \, meV$ from resonance and a rate that is below experimental error for negative detunings. Furthermore, the additional tunneling rate is appreciable for a wide range of detunings up to $+6 \, meV$, much larger than the intrinsic resonant tunneling rate that is determined by $V_1$ ($\sim +1.5 \, meV$).  

This characteristic energy scale of a few millielectronvolts corresponds to the energy of acoustic phonons in GaAs with a wavelength comparable to characteristic dimensions in the QD-molecule (width, dot-separation) \cite{Gawarecki10}. Theoretical calculations of the coupling between molecular states and phonons in similar structures taking into account deformation- and piezoelectric- exciton-phonon coupling reported a spectral density which shows a maximum around $+1.4 \,meV$ with a rate of $0.3 \, ps^{-1}$ \cite{Gawarecki10}. We interpret the additional relaxation channel as reflecting phonon mediated inelastic tunneling between the two dots forming the molecule, as schematically depicted in the inset of Fig.4(c). For positive detunings of $e_{ld}^*$ the electron tunnels into $e_{ld}^*$ whilst simultaneously emitting a phonon with an energy equal to the level separation. From there it most likely tunnels out of the excited orbital state of the QDM instead of relaxing to the ground state since the tunneling escape time from the excited state is $<5 \, ps$ (see red line in Fig.4(a)). In contrast, for negative detunings where phonon absorption would be required, such an inelastic tunneling is prohibited by the negligible phonon occupation cryogenic temperatures (see supplementary material)
%To further corroborate this interpretation we performed measurements of $t_e$ at an elevated temperature of $T=50K$. For the case of phonon emission ($e_{ld}^*$ red detuned compared with $e_{ud}$) $t_e$ remained the same for all measurements. In contrast, for the case of phonon absorption ($e_{ld}^*$ blue detuned) we typically observe a strong decrease of $t_e$ \cite{temp}. This strongly supports the model of phonon-mediated inelastic tunneling to the excited orbital state of the lower dot.

In summary, we directly investigated electron tunneling dynamics in a QDM using ultrafast pump probe spectroscopy with PC readout. We determined the electron and hole tunneling times as a function of $F$ and showed that $t_e$ exhibits resonances when resonant tunelling can occur between the two dots forming the molecule. In this regime, the tunnel coupled energy level serves as an intermediate state facilitating electron escape within a few ps. When the orbital states are detuned by a few meV, inelastic tunneling involving the emission of acoustic phonons is found to dominate the charge transfer dynamics. The results obtained demonstrate that optically generated excitons can be spatially separated within the quantum dot molecule over timescales faster than $5 \,ps$, allowing the nanostructure to be used as an optically pumped ultrafast charge memory.  

We gratefully acknowledge financial support of the DFG via SFB-631, Nanosystems Initiative Munich and the Emmy Noether Program (H.J.K.), the EU via SOLID and the TUM Graduate School. We thank J. M. Daniels, T. Kuhn and P. Machnikowski for discussions.


\begin{thebibliography}{18}%
\makeatletter
\providecommand \@ifxundefined [1]{%
 \@ifx{#1\undefined}
}%
\providecommand \@ifnum [1]{%
 \ifnum #1\expandafter \@firstoftwo
 \else \expandafter \@secondoftwo
 \fi
}%
\providecommand \@ifx [1]{%
 \ifx #1\expandafter \@firstoftwo
 \else \expandafter \@secondoftwo
 \fi
}%
\providecommand \natexlab [1]{#1}%
\providecommand \enquote  [1]{``#1''}%
\providecommand \bibnamefont  [1]{#1}%
\providecommand \bibfnamefont [1]{#1}%
\providecommand \citenamefont [1]{#1}%
\providecommand \href@noop [0]{\@secondoftwo}%
\providecommand \href [0]{\begingroup \@sanitize@url \@href}%
\providecommand \@href[1]{\@@startlink{#1}\@@href}%
\providecommand \@@href[1]{\endgroup#1\@@endlink}%
\providecommand \@sanitize@url [0]{\catcode `\\12\catcode `\$12\catcode
  `\&12\catcode `\#12\catcode `\^12\catcode `\_12\catcode `\%12\relax}%
\providecommand \@@startlink[1]{}%
\providecommand \@@endlink[0]{}%
\providecommand \url  [0]{\begingroup\@sanitize@url \@url }%
\providecommand \@url [1]{\endgroup\@href {#1}{\urlprefix }}%
\providecommand \urlprefix  [0]{URL }%
\providecommand \Eprint [0]{\href }%
\@ifxundefined \urlstyle {%
  \providecommand \doi  [0]{\begingroup \@sanitize@url \@doi}%
  \providecommand \@doi [1]{\endgroup \@@startlink {\doibase
  #1}doi:\discretionary {}{}{}#1\@@endlink }%
}{%
  \providecommand \doi  [0]{doi:\discretionary{}{}{}\begingroup
  \urlstyle{rm}\Url }%
}%
\providecommand \doibase [0]{http://dx.doi.org/}%
\providecommand \Doi [0]{\begingroup \@sanitize@url \@Doi }%
\providecommand \@Doi  [1]{\endgroup\@@startlink{\doibase#1}\@@Doi}%
\providecommand \@@Doi [1]{#1\@@endlink}%
\providecommand \selectlanguage [0]{\@gobble}%
\providecommand \bibinfo  [0]{\@secondoftwo}%
\providecommand \bibfield  [0]{\@secondoftwo}%
\providecommand \translation [1]{[#1]}%
\providecommand \BibitemOpen [0]{}%
\providecommand \bibitemStop [0]{}%
\providecommand \bibitemNoStop [0]{.\EOS\space}%
\providecommand \EOS [0]{\spacefactor3000\relax}%
\providecommand \BibitemShut  [1]{\csname bibitem#1\endcsname}%
%</preamble>
\bibitem [{\citenamefont {De~Franceschi}\ \emph {et~al.}(2001)\citenamefont
  {De~Franceschi}, \citenamefont {Sasaki}, \citenamefont {Elzerman},
  \citenamefont {van~der Wiel}, \citenamefont {Tarucha},\ and\ \citenamefont
  {Kouwenhoven}}]{PhysRevLett.86.878}%
  \BibitemOpen
  \bibfield  {author} {\bibinfo {author} {\bibfnamefont {S.}~\bibnamefont
  {De~Franceschi}}, \bibinfo {author} {\bibfnamefont {S.}~\bibnamefont
  {Sasaki}}, \bibinfo {author} {\bibfnamefont {J.~M.}\ \bibnamefont
  {Elzerman}}, \bibinfo {author} {\bibfnamefont {W.~G.}\ \bibnamefont {van~der
  Wiel}}, \bibinfo {author} {\bibfnamefont {S.}~\bibnamefont {Tarucha}}, \ and\
  \bibinfo {author} {\bibfnamefont {L.~P.}\ \bibnamefont {Kouwenhoven}},\ }\Doi
  {10.1103/PhysRevLett.86.878} {\bibfield  {journal} {\bibinfo  {journal}
  {Phys. Rev. Lett.},\ }\textbf {\bibinfo {volume} {86}},\ \bibinfo {pages}
  {878} (\bibinfo {year} {2001})}\BibitemShut {NoStop}%
\bibitem [{\citenamefont {Qin}\ \emph {et~al.}(2001)\citenamefont {Qin},
  \citenamefont {Holleitner}, \citenamefont {Eberl},\ and\ \citenamefont
  {Blick}}]{Qin01}%
  \BibitemOpen
  \bibfield  {author} {\bibinfo {author} {\bibfnamefont {H.}~\bibnamefont
  {Qin}}, \bibinfo {author} {\bibfnamefont {A.}~\bibnamefont {Holleitner}},
  \bibinfo {author} {\bibfnamefont {K.}~\bibnamefont {Eberl}}, \ and\ \bibinfo
  {author} {\bibfnamefont {R.}~\bibnamefont {Blick}},\ }\Doi
  {{10.1103/PhysRevB.64.241302}} {\bibfield  {journal} {\bibinfo  {journal}
  {{Phys. Rev. B}},\ }\textbf {\bibinfo {volume} {{64}}} (\bibinfo {year}
  {{2001}})},\ ISSN \bibinfo {issn} {{1098-0121}},\ \doi
  {{10.1103/PhysRevB.64.241302}}\BibitemShut {NoStop}%
\bibitem [{\citenamefont {Mason}\ \emph {et~al.}(2004)\citenamefont {Mason},
  \citenamefont {Biercuk},\ and\ \citenamefont {Marcus}}]{Mason30012004}%
  \BibitemOpen
  \bibfield  {author} {\bibinfo {author} {\bibfnamefont {N.}~\bibnamefont
  {Mason}}, \bibinfo {author} {\bibfnamefont {M.~J.}\ \bibnamefont {Biercuk}},
  \ and\ \bibinfo {author} {\bibfnamefont {C.~M.}\ \bibnamefont {Marcus}},\
  }\Doi {10.1126/science.1093605} {\bibfield  {journal} {\bibinfo  {journal}
  {Science},\ }\textbf {\bibinfo {volume} {303}},\ \bibinfo {pages} {655}
  (\bibinfo {year} {2004})},\ \Eprint
  {http://arxiv.org/abs/http://www.sciencemag.org/content/303/5658/655.full.pd%
f} {http://www.sciencemag.org/content/303/5658/655.full.pdf} \BibitemShut
  {NoStop}%
\bibitem [{\citenamefont {Krenner}\ \emph {et~al.}(2005)\citenamefont
  {Krenner}, \citenamefont {Sabathil}, \citenamefont {Clark}, \citenamefont
  {Kress}, \citenamefont {Schuh}, \citenamefont {Bichler}, \citenamefont
  {Abstreiter},\ and\ \citenamefont {Finley}}]{PhysRevLett.94.057402}%
  \BibitemOpen
  \bibfield  {author} {\bibinfo {author} {\bibfnamefont {H.~J.}\ \bibnamefont
  {Krenner}}, \bibinfo {author} {\bibfnamefont {M.}~\bibnamefont {Sabathil}},
  \bibinfo {author} {\bibfnamefont {E.~C.}\ \bibnamefont {Clark}}, \bibinfo
  {author} {\bibfnamefont {A.}~\bibnamefont {Kress}}, \bibinfo {author}
  {\bibfnamefont {D.}~\bibnamefont {Schuh}}, \bibinfo {author} {\bibfnamefont
  {M.}~\bibnamefont {Bichler}}, \bibinfo {author} {\bibfnamefont
  {G.}~\bibnamefont {Abstreiter}}, \ and\ \bibinfo {author} {\bibfnamefont
  {J.~J.}\ \bibnamefont {Finley}},\ }\Doi {10.1103/PhysRevLett.94.057402}
  {\bibfield  {journal} {\bibinfo  {journal} {Phys. Rev. Lett.},\ }\textbf
  {\bibinfo {volume} {94}},\ \bibinfo {pages} {057402} (\bibinfo {year}
  {2005})}\BibitemShut {NoStop}%
\bibitem [{\citenamefont {Stinaff}\ \emph {et~al.}(2006)\citenamefont
  {Stinaff}, \citenamefont {Scheibner}, \citenamefont {Bracker}, \citenamefont
  {Ponomarev}, \citenamefont {Korenev}, \citenamefont {Ware}, \citenamefont
  {Doty}, \citenamefont {Reinecke},\ and\ \citenamefont
  {Gammon}}]{Stinaff2006}%
  \BibitemOpen
  \bibfield  {author} {\bibinfo {author} {\bibfnamefont {E.}~\bibnamefont
  {Stinaff}}, \bibinfo {author} {\bibfnamefont {M.}~\bibnamefont {Scheibner}},
  \bibinfo {author} {\bibfnamefont {A.}~\bibnamefont {Bracker}}, \bibinfo
  {author} {\bibfnamefont {I.}~\bibnamefont {Ponomarev}}, \bibinfo {author}
  {\bibfnamefont {V.}~\bibnamefont {Korenev}}, \bibinfo {author} {\bibfnamefont
  {M.}~\bibnamefont {Ware}}, \bibinfo {author} {\bibfnamefont {M.}~\bibnamefont
  {Doty}}, \bibinfo {author} {\bibfnamefont {T.}~\bibnamefont {Reinecke}}, \
  and\ \bibinfo {author} {\bibfnamefont {D.}~\bibnamefont {Gammon}},\
  }\href@noop {} {\bibfield  {journal} {\bibinfo  {journal} {Science},\
  }\textbf {\bibinfo {volume} {311}},\ \bibinfo {pages} {636} (\bibinfo {year}
  {2006})}\BibitemShut {NoStop}%
\bibitem [{\citenamefont {Krenner}\ \emph {et~al.}(2006)\citenamefont
  {Krenner}, \citenamefont {Clark}, \citenamefont {Nakaoka}, \citenamefont
  {Bichler}, \citenamefont {Scheurer}, \citenamefont {Abstreiter},\ and\
  \citenamefont {Finley}}]{PhysRevLett.97.076403}%
  \BibitemOpen
  \bibfield  {author} {\bibinfo {author} {\bibfnamefont {H.~J.}\ \bibnamefont
  {Krenner}}, \bibinfo {author} {\bibfnamefont {E.~C.}\ \bibnamefont {Clark}},
  \bibinfo {author} {\bibfnamefont {T.}~\bibnamefont {Nakaoka}}, \bibinfo
  {author} {\bibfnamefont {M.}~\bibnamefont {Bichler}}, \bibinfo {author}
  {\bibfnamefont {C.}~\bibnamefont {Scheurer}}, \bibinfo {author}
  {\bibfnamefont {G.}~\bibnamefont {Abstreiter}}, \ and\ \bibinfo {author}
  {\bibfnamefont {J.~J.}\ \bibnamefont {Finley}},\ }\Doi
  {10.1103/PhysRevLett.97.076403} {\bibfield  {journal} {\bibinfo  {journal}
  {Phys. Rev. Lett.},\ }\textbf {\bibinfo {volume} {97}},\ \bibinfo {pages}
  {076403} (\bibinfo {year} {2006})}\BibitemShut {NoStop}%
\bibitem [{\citenamefont {Bracker}\ \emph {et~al.}(2006)\citenamefont
  {Bracker}, \citenamefont {Scheibner}, \citenamefont {Doty}, \citenamefont
  {Stinaff}, \citenamefont {Ponomarev}, \citenamefont {Kim}, \citenamefont
  {Whitman}, \citenamefont {Reinecke},\ and\ \citenamefont
  {Gammon}}]{Bracker2006}%
  \BibitemOpen
  \bibfield  {author} {\bibinfo {author} {\bibfnamefont {A.~S.}\ \bibnamefont
  {Bracker}}, \bibinfo {author} {\bibfnamefont {M.}~\bibnamefont {Scheibner}},
  \bibinfo {author} {\bibfnamefont {M.~F.}\ \bibnamefont {Doty}}, \bibinfo
  {author} {\bibfnamefont {E.~A.}\ \bibnamefont {Stinaff}}, \bibinfo {author}
  {\bibfnamefont {I.~V.}\ \bibnamefont {Ponomarev}}, \bibinfo {author}
  {\bibfnamefont {J.~C.}\ \bibnamefont {Kim}}, \bibinfo {author} {\bibfnamefont
  {L.~J.}\ \bibnamefont {Whitman}}, \bibinfo {author} {\bibfnamefont {T.~L.}\
  \bibnamefont {Reinecke}}, \ and\ \bibinfo {author} {\bibfnamefont
  {D.}~\bibnamefont {Gammon}},\ }\Doi {10.1063/1.2400397} {\bibfield  {journal}
  {\bibinfo  {journal} {Appl. Phys. Lett.},\ }\textbf {\bibinfo {volume}
  {89}},\ \bibinfo {pages} {233110} (\bibinfo {year} {2006})},\ ISSN \bibinfo
  {issn} {0003-6951}\BibitemShut {NoStop}%
\bibitem [{\citenamefont {Scheibner}\ \emph {et~al.}(2008)\citenamefont
  {Scheibner}, \citenamefont {Yakes}, \citenamefont {Bracker}, \citenamefont
  {Ponomarev}, \citenamefont {Doty}, \citenamefont {Hellberg}, \citenamefont
  {Whitman}, \citenamefont {Reinecke},\ and\ \citenamefont
  {Gammon}}]{Scheibner2008}%
  \BibitemOpen
  \bibfield  {author} {\bibinfo {author} {\bibfnamefont {M.}~\bibnamefont
  {Scheibner}}, \bibinfo {author} {\bibfnamefont {M.}~\bibnamefont {Yakes}},
  \bibinfo {author} {\bibfnamefont {A.~S.}\ \bibnamefont {Bracker}}, \bibinfo
  {author} {\bibfnamefont {I.~V.}\ \bibnamefont {Ponomarev}}, \bibinfo {author}
  {\bibfnamefont {M.~F.}\ \bibnamefont {Doty}}, \bibinfo {author}
  {\bibfnamefont {C.~S.}\ \bibnamefont {Hellberg}}, \bibinfo {author}
  {\bibfnamefont {L.~J.}\ \bibnamefont {Whitman}}, \bibinfo {author}
  {\bibfnamefont {T.~L.}\ \bibnamefont {Reinecke}}, \ and\ \bibinfo {author}
  {\bibfnamefont {D.}~\bibnamefont {Gammon}},\ }\Doi {10.1038/nphys882}
  {\bibfield  {journal} {\bibinfo  {journal} {Nature Physics},\ }\textbf
  {\bibinfo {volume} {4}},\ \bibinfo {pages} {291} (\bibinfo {year} {2008})},\
  ISSN \bibinfo {issn} {1745-2473}\BibitemShut {NoStop}%
\bibitem [{\citenamefont {M\"uller}\ \emph {et~al.}(2011)\citenamefont
  {M\"uller}, \citenamefont {Reithmaier}, \citenamefont {Clark}, \citenamefont
  {Jovanov}, \citenamefont {Bichler}, \citenamefont {Krenner}, \citenamefont
  {Betz}, \citenamefont {Abstreiter},\ and\ \citenamefont
  {Finley}}]{Mueller11}%
  \BibitemOpen
  \bibfield  {author} {\bibinfo {author} {\bibfnamefont {K.}~\bibnamefont
  {M\"uller}}, \bibinfo {author} {\bibfnamefont {G.}~\bibnamefont
  {Reithmaier}}, \bibinfo {author} {\bibfnamefont {E.~C.}\ \bibnamefont
  {Clark}}, \bibinfo {author} {\bibfnamefont {V.}~\bibnamefont {Jovanov}},
  \bibinfo {author} {\bibfnamefont {M.}~\bibnamefont {Bichler}}, \bibinfo
  {author} {\bibfnamefont {H.~J.}\ \bibnamefont {Krenner}}, \bibinfo {author}
  {\bibfnamefont {M.}~\bibnamefont {Betz}}, \bibinfo {author} {\bibfnamefont
  {G.}~\bibnamefont {Abstreiter}}, \ and\ \bibinfo {author} {\bibfnamefont
  {J.~J.}\ \bibnamefont {Finley}},\ }\Doi {10.1103/PhysRevB.84.081302}
  {\bibfield  {journal} {\bibinfo  {journal} {Phys. Rev. B},\ }\textbf
  {\bibinfo {volume} {84}},\ \bibinfo {pages} {081302} (\bibinfo {year}
  {2011})}\BibitemShut {NoStop}%
\bibitem [{com()}]{complex}%
  \BibitemOpen
  \href@noop {} {}\bibinfo {note} {The additional transitions observed in Fig.
  1 arise primarily from charged excitons [see e.g. ref. 6] due to the
  statistically fluctuating occupancy of the QD-molecule by photogenerated
  charge carriers. Such transitions are not probed in PC-experiments since free
  carriers are not generated in the vicinity of the QD-molecule following
  strictly resonant excitation}\BibitemShut {NoStop}%
\bibitem [{par()}]{parity}%
  \BibitemOpen
  \href@noop {} {}\bibinfo {note} {The resulting coupling is $V_1/V_0 = 4.25$
  times weaker than the principle anticrossing at $F_0$ since the participating
  electronic orbitals have different parity ; s-like for $e_{ud}$ and p-like
  for $e_{ld}^*$}\BibitemShut {NoStop}%
\bibitem [{\citenamefont {Zecherle}\ \emph {et~al.}(2010)\citenamefont
  {Zecherle}, \citenamefont {Ruppert}, \citenamefont {Clark}, \citenamefont
  {Abstreiter}, \citenamefont {Finley},\ and\ \citenamefont
  {Betz}}]{Zecherle2010}%
  \BibitemOpen
  \bibfield  {author} {\bibinfo {author} {\bibfnamefont {M.}~\bibnamefont
  {Zecherle}}, \bibinfo {author} {\bibfnamefont {C.}~\bibnamefont {Ruppert}},
  \bibinfo {author} {\bibfnamefont {E.~C.}\ \bibnamefont {Clark}}, \bibinfo
  {author} {\bibfnamefont {G.}~\bibnamefont {Abstreiter}}, \bibinfo {author}
  {\bibfnamefont {J.~J.}\ \bibnamefont {Finley}}, \ and\ \bibinfo {author}
  {\bibfnamefont {M.}~\bibnamefont {Betz}},\ }\Doi {10.1103/PhysRevB.82.125314}
  {\bibfield  {journal} {\bibinfo  {journal} {Phys Rev B},\ }\textbf {\bibinfo
  {volume} {82}},\ \bibinfo {pages} {125314} (\bibinfo {year} {2010})},\ ISSN
  \bibinfo {issn} {1098-0121}\BibitemShut {NoStop}%
\bibitem [{\citenamefont {Ramsay}\ \emph {et~al.}(2008)\citenamefont {Ramsay},
  \citenamefont {Boyle}, \citenamefont {Kolodka}, \citenamefont {Oliveira},
  \citenamefont {Skiba-Szymanska}, \citenamefont {Liu}, \citenamefont
  {Hopkinson}, \citenamefont {Fox},\ and\ \citenamefont {Skolnick}}]{Ramsay08}%
  \BibitemOpen
  \bibfield  {author} {\bibinfo {author} {\bibfnamefont {A.~J.}\ \bibnamefont
  {Ramsay}}, \bibinfo {author} {\bibfnamefont {S.~J.}\ \bibnamefont {Boyle}},
  \bibinfo {author} {\bibfnamefont {R.~S.}\ \bibnamefont {Kolodka}}, \bibinfo
  {author} {\bibfnamefont {J.~B.~B.}\ \bibnamefont {Oliveira}}, \bibinfo
  {author} {\bibfnamefont {J.}~\bibnamefont {Skiba-Szymanska}}, \bibinfo
  {author} {\bibfnamefont {H.~Y.}\ \bibnamefont {Liu}}, \bibinfo {author}
  {\bibfnamefont {M.}~\bibnamefont {Hopkinson}}, \bibinfo {author}
  {\bibfnamefont {A.~M.}\ \bibnamefont {Fox}}, \ and\ \bibinfo {author}
  {\bibfnamefont {M.~S.}\ \bibnamefont {Skolnick}},\ }\Doi
  {{10.1103/PhysRevLett.100.197401}} {\bibfield  {journal} {\bibinfo  {journal}
  {{Phys. Rev. Lett.}},\ }\textbf {\bibinfo {volume} {{100}}} (\bibinfo {year}
  {{2008}})},\ ISSN \bibinfo {issn} {{0031-9007}},\ \doi
  {{10.1103/PhysRevLett.100.197401}}\BibitemShut {NoStop}%
\bibitem [{\citenamefont {Boyle}\ \emph {et~al.}(2008)\citenamefont {Boyle},
  \citenamefont {Ramsay}, \citenamefont {Bello}, \citenamefont {Liu},
  \citenamefont {Hopkinson}, \citenamefont {Fox},\ and\ \citenamefont
  {Skolnick}}]{Boyle08}%
  \BibitemOpen
  \bibfield  {author} {\bibinfo {author} {\bibfnamefont {S.~J.}\ \bibnamefont
  {Boyle}}, \bibinfo {author} {\bibfnamefont {A.~J.}\ \bibnamefont {Ramsay}},
  \bibinfo {author} {\bibfnamefont {F.}~\bibnamefont {Bello}}, \bibinfo
  {author} {\bibfnamefont {H.~Y.}\ \bibnamefont {Liu}}, \bibinfo {author}
  {\bibfnamefont {M.}~\bibnamefont {Hopkinson}}, \bibinfo {author}
  {\bibfnamefont {A.~M.}\ \bibnamefont {Fox}}, \ and\ \bibinfo {author}
  {\bibfnamefont {M.~S.}\ \bibnamefont {Skolnick}},\ }\Doi
  {{10.1103/PhysRevB.78.075301}} {\bibfield  {journal} {\bibinfo  {journal}
  {{Phys. Rev. B}},\ }\textbf {\bibinfo {volume} {{78}}} (\bibinfo {year}
  {{2008}})},\ ISSN \bibinfo {issn} {{1098-0121}},\ \doi
  {{10.1103/PhysRevB.78.075301}}\BibitemShut {NoStop}%
\bibitem [{\citenamefont {Stufler}\ \emph {et~al.}(2006)\citenamefont
  {Stufler}, \citenamefont {Machnikowski}, \citenamefont {Ester}, \citenamefont
  {Bichler}, \citenamefont {Axt}, \citenamefont {Kuhn},\ and\ \citenamefont
  {Zrenner}}]{PhysRevB.73.125304}%
  \BibitemOpen
  \bibfield  {author} {\bibinfo {author} {\bibfnamefont {S.}~\bibnamefont
  {Stufler}}, \bibinfo {author} {\bibfnamefont {P.}~\bibnamefont
  {Machnikowski}}, \bibinfo {author} {\bibfnamefont {P.}~\bibnamefont {Ester}},
  \bibinfo {author} {\bibfnamefont {M.}~\bibnamefont {Bichler}}, \bibinfo
  {author} {\bibfnamefont {V.~M.}\ \bibnamefont {Axt}}, \bibinfo {author}
  {\bibfnamefont {T.}~\bibnamefont {Kuhn}}, \ and\ \bibinfo {author}
  {\bibfnamefont {A.}~\bibnamefont {Zrenner}},\ }\Doi
  {10.1103/PhysRevB.73.125304} {\bibfield  {journal} {\bibinfo  {journal}
  {Phys. Rev. B},\ }\textbf {\bibinfo {volume} {73}},\ \bibinfo {pages}
  {125304} (\bibinfo {year} {2006})}\BibitemShut {NoStop}%
\bibitem [{\citenamefont {Villas-Boas}\ \emph {et~al.}(2005)\citenamefont
  {Villas-Boas}, \citenamefont {Ulloa},\ and\ \citenamefont
  {Govorov}}]{Villas-Boas2005}%
  \BibitemOpen
  \bibfield  {author} {\bibinfo {author} {\bibfnamefont {J.}~\bibnamefont
  {Villas-Boas}}, \bibinfo {author} {\bibfnamefont {S.}~\bibnamefont {Ulloa}},
  \ and\ \bibinfo {author} {\bibfnamefont {A.}~\bibnamefont {Govorov}},\ }\Doi
  {{10.1103/PhysRevLett.94.057404}} {\bibfield  {journal} {\bibinfo  {journal}
  {{Phys. Rev. Lett.}},\ }\textbf {\bibinfo {volume} {{94}}} (\bibinfo {year}
  {{2005}})},\ ISSN \bibinfo {issn} {{0031-9007}},\ \doi
  {{10.1103/PhysRevLett.94.057404}}\BibitemShut {NoStop}%
\bibitem [{\citenamefont {Gawarecki}\ \emph {et~al.}(2010)\citenamefont
  {Gawarecki}, \citenamefont {Pochwa\l{}a}, \citenamefont
  {Grodecka\char21{}Grad},\ and\ \citenamefont {Machnikowski}}]{Gawarecki10}%
  \BibitemOpen
  \bibfield  {author} {\bibinfo {author} {\bibfnamefont {K.}~\bibnamefont
  {Gawarecki}}, \bibinfo {author} {\bibfnamefont {M.}~\bibnamefont
  {Pochwa\l{}a}}, \bibinfo {author} {\bibfnamefont {A.}~\bibnamefont
  {Grodecka\char21{}Grad}}, \ and\ \bibinfo {author} {\bibfnamefont
  {P.}~\bibnamefont {Machnikowski}},\ }\Doi {10.1103/PhysRevB.81.245312}
  {\bibfield  {journal} {\bibinfo  {journal} {Phys. Rev. B},\ }\textbf
  {\bibinfo {volume} {81}},\ \bibinfo {pages} {245312} (\bibinfo {year}
  {2010})}\BibitemShut {NoStop}%
\bibitem [{tem()}]{temp}%
  \BibitemOpen
  \href@noop {} {}\bibinfo {note} {For example for a detuning of $\Delta E =
  -1.33meV$ we observe a reduction of $t_e$ from $42.5 \pm 2 \, ps$ to $26 \pm
  2 \, ps$ and at $\Delta E = -2.4meV$ from $58 \pm 2 \, ps$ to $34 \pm 3 \,
  ps$}\BibitemShut {NoStop}%
\end{thebibliography}
\end{document}